\documentclass[aps,preprint,floatfix,nofootinbib,showpacs]{revtex4-1}
\pdfoutput=1
\usepackage{graphicx,color,float, amsmath}
\usepackage{hyperref}
\usepackage{MnSymbol}%
\usepackage{wasysym}%
\usepackage{color}
\usepackage{slashed}

\begin{document}

{\small
\begin{flushright}
\end{flushright} }

\title{ Effects of Kaluza-Klein Neutrinos on $R_{D}$ and $R_{D^{*}}$}
\vspace*{1cm}

\author{
Janus Capellan Aban, Chuan-Ren Chen, Chrisna~Setyo~Nugroho}
\affiliation{
\vspace*{.5cm}
Department of Physics, National Taiwan Normal University, Taipei 116, Taiwan 
\vspace*{1cm}}

\begin{abstract}
Recent measurements of $R_{D}$ and $R_{D^{*}}$ by the LHCb
collaboration show  deviations from their respective Standard
Model values. These semileptonic $B$ meson decays, associated
with $b\rightarrow c \tau \bar{\nu}$ transition, are pointing toward
new physics beyond the Standard Model via leptonic flavor universality violation. In this paper, we show that such anomaly can be resolved by
the cummulative Kaluza-Klein (KK) modes of singlet right-handed
neutrino which propagates in the large extra dimensional space. We found that the number of extra dimension should be 2 to explain $R_{D}$ and $R_{D^{*}}$.   We
show that both $R_{D}$ and $R_{D^{*}}$ constraint the energy scale $M_{F}$ of this extra dimension which are compatible with the limits
from lepton flavor violating tau decays. In contrast, our
findings are in tension with the limits coming from the neutrino
experiments which set the most stringent lower bound on $M_{F}$.
The future measurements of $R_{D^{(*)}}^{exp}$  with reduced uncertainties will exclude this extra dimensional model with right-handed neutrino propagating in the bulk, if the central values stay.
\end{abstract}

\maketitle

\section{Introduction}
Discrepancies between the Standard Model (SM) predictions and experimental data in the decays of B mesons have gained much attention for years since they may be the hints of new physics beyond the SM, in particular the violations of lepton flavor universality (LFU) in the measurements of $R_{K^{(*)}}$ and $R_{D^{(*)}}$.  Even though the updated LHCb measurement~\cite{LHCb:2022qnv} confirmed the consistency with SM predictions in $R_K$ and $R_{K^{*}}$ due to the $b\to s$ transitions, the LFU violation remains about $3\sigma$ away from the SM expectations in $b\rightarrow c \tau \bar{\nu}_\tau$ transition in the  $R_D$ and $R_{D^{*}}$ measurements.

The definition of $R_{D^{(*)}}$ is given as 
\begin{equation}
R_{D^{(*)}}=\dfrac{Br(B\rightarrow D^{(*)}\tau \bar{\nu}_{\tau} )}{[(Br(B\rightarrow D^{(*)}e \bar{\nu}_{e} + Br(B\rightarrow D^{(*)}\mu \bar{\nu}_{\mu} ))/2]},
\end{equation}
which is independent of $|V_{cb}|$ and also of the $B\to D^{(*)}$ form factor to a large extent~\cite{HFLAV:2022pwe}.
A newly calculated world average~\cite{Iguro} using the data from Belle, BaBar, and LHCb collaborations in 2022 obtained $R_{D}^{ave}=0.358\pm 0.027$ and $R_{D^*}^{ave}=0.285\pm 0.013$. The SM predictions of the branching ratios $R_{D}^{SM}$ and $R_{D^*}^{SM}$ are $0.298\pm 0.004$ and $0.254 \pm 0.005$~\cite{HFLAV:2022pwe}, respectively, which are clearly smaller than the measurements. After incorporating all the recent developments in $B\rightarrow D^{(*)}$ that include form factors for predicting $R_{D^{(*)}}^{SM}$, the largest pull of the combined new world average is $4.1\sigma$ from SM predictions~\cite{Iguro}. Furthermore, even the most updated results of LHCb~\cite{HFLAV:2023prelim} using the data collected in 2015 and 2016 are included, the global picture of the combined new world average does not change. And this corresponds to the most recent combined world average $R_{D}^{ave}=0.356\pm 0.029$ and $R_{D^*}^{ave}=0.284\pm 0.013$.
For this large deviation, the new physics effect could be comparable to the tree-level SM contributions, therefore many models have been proposed, such as introducing leptoquarks~\cite{Hiller:2021pul}, or new colorless vector $W'$~\cite{Megias:2017ove}, or scalar particles~\cite{Tanaka:1994ay} as tree-level mediators. 

In this work, we study a large extra-dimensional model in which three generations of right-handed neutrinos propagate in the bulk~\cite{Strumia:2000}. As a result of the compactification of these fields, active neutrinos become massive and eventually mix with KK neutrinos due to the mass matrix diagonalization process. Concerning $R_{D}^{(*)}$, it is important to note that SM tree-level semileptonic processes with $W^{\pm}$ as mediators preserved the LFU in the SM. 
With the existence of KK neutrinos, there may be additional decay channels for $b$ decays into $c$ and KK neutrinos, $b\rightarrow c \ell \bar{\nu}_{\ell}^{KK}$, if KK neutrinos are light enough. Furthermore, if $b\rightarrow c \tau \bar{\nu}_{\tau}^{KK}$ decay width is much lager than  $b\rightarrow c e \bar{\nu}_{e}^{KK}$  and $b\rightarrow c \mu \bar{\nu}_{\mu}^{KK}$, the $R_{D^{(*)}}$ would be larger than the SM prediction, and this is the case we consider in this study. Namely, 
new physics contributes to $R_{D}^{(*)}$ from the cumulative effects of $\tau$ neutrino KK modes in $b\rightarrow c \tau \bar{\nu}_{\tau}^{KK}$ transition as a tree-level process via $W^{\pm}$ exchange.

The following is how the paper is structured: we briefly discuss the extra-dimensional KK model described by~\cite{Strumia:2000} in section~\ref{sec:model}. Then in section~\ref{sec:btoctaunu}, we calculate the decay width $\Gamma (B\rightarrow D \tau \bar{\nu}_{\tau}^{KK})$ in a tree-level process to determine the total contributions of the KK neutrinos to $R_{D^{(*)}}$ measurements.
Section~\ref{sec:constraints} discusses the constraints we consider for the lower limits of fundamental scale $M_F$ of extra dimension. In addition to this part, the results of a combined analysis of several neutrino experiments performed in~\cite{Forero:2022skg} to confront the upper bound of the Large Extra Dimension (LED) size $R$ has been implemented. 
From~\cite{Forero:2022skg}, the upper bound for  $R$ at $90\%$ confidence level (C.L.) is $R<0.20 \,\mu m$ for normal ordering (NO) and $R<0.10 \,\mu m $ for inverted ordering (IO). Finally, in section~\ref{sec:con} we present our conclusions.

\section{Model}
 \label{sec:model}
In this section, we briefly review a model in the extra-dimensional framework, where three right-handed neutrinos are introduced and able to propagate in the bulk, while all the SM particles are on the brane~\cite{Strumia:2000}.   

The effective action of such interaction is given as~\cite{Strumia:2000}
\begin{align}
\label{eq: action1}
S=\int d^4x dy [ \bar{\Psi_i} \Gamma_A i \partial^A \Psi_i] + \int d^4x[\bar{\nu_i}i\slashed{\partial}\nu_i + \nu_i \lambda_{ij} \psi_{j}(x^\mu, 0)H + h.c.]\,,
\end{align}
where $A=0,1,2,3,4$, $\Psi_i(x,y )$ and $\nu_i$ with $i=1,2,3$ are respectively the right-handed neutrino and active neutrino fields, $H$ is the 4D Higgs doublet, and $\lambda$ is the matrix of Yukawa couplings. The right-handed neutrino fields can be written as 
\begin{align}
\label{eq:bi-spinor}
\Psi_j(x,y)=
\begin{pmatrix}
\psi_{j}(x,y) \\
\bar{\psi}_{j}^{c}(x,y)
\end{pmatrix}\,,
\end{align}
such that  $\psi_j$ and $\bar{\psi}_j^c$ are two-component Weyl spinors in five dimensions. Here, we denote $x\equiv x^\mu$ as the four-vector with $\mu=0,1,2,3$, where $x^0$ is the time coordinate and $x^i$ are the spatial coordinates for each $i=1,2,3$. The extra-dimensional coordinates are represented by  $y\equiv y^k$, where $k=1,2,...,\delta$ with $\delta$ the number of extra dimensions. Note that the fundamental scale $M_F$ is related to the Planck scale $M_p\simeq 10^{19}$ GeV by $M_F^{\delta+2}\simeq M_{P}^{2} R^{-\delta}$.
Suppose $\Psi_j(x,y)$ are $2\pi R$-periodic on variable $y$ then we can express its components into Fourier modes as 
\begin{align}
\label{eq:1stCom}
&\psi_{j}(x,y) = \frac{1}{\sqrt{2\pi R}} \sum_{n=-\infty}^{+\infty} \psi_{j}^{(n)} (x)
\exp(\frac{iny}{R})\,, \\
&\psi_{j}^{c}(x,y) = \frac{1}{\sqrt{2\pi R}} \sum_{n=-\infty}^{+\infty} \psi_{j}^{(n)c}(x) \exp(\frac{iny}{R})\,. 
\end{align}

 With a redefinition of the standard left-handed neutrinos $\nu_i$ with its neutrino flavor eigenstates, the relation

\begin{align}
\label{eq:Vneutrino}
\nu_{\alpha,L}^{f}=\sum_{i=1,2,3} V_{\alpha i}\nu_i, \:\:\:\:\:\:\:\: \alpha=e, \mu, \tau.
\end{align} 

and the compactification in coordinate $y$ on a circle of radius $R$, the action ~\eqref{eq: action1}, after spontaneous symmetry breaking, will give $S=\int \mathcal{L}d^4x$~\cite{Strumia:2000}, where
\begin{align}
\label{eq:Lagrangians}
\mathcal{L}=\sum_{i=1,2,3} \left( \bar{\nu}_i i\slashed{\partial}\nu_i + \sum_{n=-\infty}^{+\infty}\Big[
\bar{\psi}_{i}^{(n)} i \slashed{\partial}\psi_{i}^{(n)} + \bar{\psi}_{i}^{(n)c } i \slashed{\partial}\psi_{i}^{(n)c} + m_i \nu_i \psi_{i}^{(n)}
\Big] + \sum_{n=1}^{+\infty} \dfrac{n}{R} (\psi_{i}^{(n)} \psi_{i}^{(n)c} - \psi_{i}^{(-n)} \psi_{i}^{(-n)c} ) + h.c. \right)
\end{align}
such that $m_i=\dfrac{M_F}{M_P} \dfrac{h_i v }{\sqrt{2}}$ as in~\cite{Ioannisian:1999cw}, with $h_i$ as the corresponding Yukawa coupling for $i=e,\mu, \tau$. It is important to note that $m_i$ is suppressed by a volume factor $\dfrac{M_F}{M_P}$ of the extra compactified dimensions~\cite{Hamed, Dienes}. Finally the relevant mass terms in the action is given by
\begin{align}
\label{eq:Lagrangianmass}
\mathcal{L}_{mass}^{KK}=\sum_{i,j=1}^{3} N_{Li}^{T} M_{ij} N_{Rj} + h.c.\,, 
\end{align} 
with KK index being suppressed, and the mass matrix is given by
\begin{align}
\label{eq:massmatrix}
M =\begin{pmatrix} 
 m_{i}& \sqrt{2} m_{i}& \sqrt{2} m_{i}& ...\\
	 0& \frac{1}{R}& 0& ... \\
	 0 & 0& \frac{2}{R}& ... \\
	\vdots& \vdots& \vdots& \ddots\\
\end{pmatrix}\,.
\end{align}
The corresponding basis vectors for this mass matrix are
\begin{align}
\label{eq:Lagrangianmass}
N_{Li}=\begin{pmatrix} 
 \nu \\
	 \nu_{nL} \\
\end{pmatrix}_i, 
\:\:\:\:\: \textrm{and}\:\:\:\:\:
N_{Ri}=\begin{pmatrix} 
 \nu_R \\
	 \nu_{nR} \\
\end{pmatrix}_i\,, 
\end{align}
where we have redefined the fields as \cite{Strumia:2000} 
\begin{align}
\label{eq:redefinitionfields}
\nu_{nR. i}=\dfrac{1}{\sqrt{2}}(\psi_{i}^{(n)}+\psi_{i}^{(-n)} ), \:\:\:\:\:\nu_{(n)L, i}=\dfrac{1}{\sqrt{2}}(\psi_{i}^{(n)c}+\psi_{i}^{(-n)c} ) \:\:\:\:\: \textrm{for}\:\:\:\:\: n\geq 1 \:\:\:\:\:\textrm{and}\:\:\:\: \nu_R=\psi_{i}^{0}.
\end{align}
We diagonalize the matrix $MM^{T}$ for each generation $i$ using a unitary matrix $U^{(i)}$ to obtain the square of the masses of the mass eigenstates. Let's call the eigenvalues of  $MM^{T}$ to be $\dfrac{\lambda_{i}^{(n)2}}{R^2}$ which satisfies~\cite{Strumia:2000}
\begin{align}
\label{eq:eigenvalueequation}
\lambda_{i}^{(n)2}-\pi\lambda_{i}^{(n)}\xi_i^2\cot \pi\lambda_{i}^{(n)}=0\,,
\end{align}
for $n\geq 0$ and $\xi_i=m_i R$ with $i=e,~\mu,~\tau$. 

Following~\cite{Ioannisian:1999cw,Langacker} the mixing component of KK neutrinos for active neutrinos can be written as
\begin{align}
\label{eq:mixingexpression1}
\Xi^{(i)}\equiv \big( \dfrac{U_{01}^{(i)}}{U_{00}^{(i)}}, \dfrac{U_{02}^{(i)}}{U_{00}^{(i)}}, \dfrac{U_{03}^{(i)}}{U_{00}^{(i)}},... \big)\,.
\end{align}
As shown in \cite{Strumia:2000, Dienes, Dvali, Mohapatra} by setting $\xi_i=m_i R$ we obtain
\begin{align}
\label{eq:mixingexpression2}
U_{0n}^{(i)2} = \dfrac{2}{1+  \pi^2  \xi_{i}^{2} + \dfrac{  \lambda_{i}^{(n)2}}{\xi_{i}^{2} }}\,.
\end{align}
In the case of $\xi_i<<1$ the approximation $\lambda_{n}^{(i)}\approx n$ for $n>0$ gives
\begin{align}
\label{eq:mixingapprox}
U_{0n}^{(i)2}\approx \dfrac{2 \xi_{i}^{2} }{ \xi_{i}^{2} +n^2}.
\end{align}
Consequently, for $n=0$, $\lambda_{0}^{(i)}$ is approximately equal to $\xi_i$ which yields $U_{00}^{(i)2}\approx 1$. Therefore it can be easily checked in equation~\eqref{eq:mixingexpression1} that
\begin{align}
\label{eq:mixingsimplified}
\Xi^{(i)}\approx ( U_{01}^{(i)}, U_{02}^{(i)}, U_{03}^{(i)}, ...).
\end{align}

The following is the relevant interaction Lagrangian involving the neutrino mass eigenstates $\nu_{l}$ and $\chi^{(n)}$, the charged leptons $l$, together with the weak bosons $W^{\pm}$, and their corresponding Goldstone bosons $G^{\pm}$ given by~\cite{Schechter}
\begin{align}
\label{eq:InteractionLagrangian}
\mathcal{L}_{W^{\pm}}^{int}&=-\dfrac{g_w}{\sqrt{2}} W^{-\mu} \sum_{l=e, \mu, \tau}
\Big( B_{l\nu_l}\bar{l}\gamma_\mu P_L \nu_l + \sum_{n=-\infty}^{\infty} 
 B_{l, n}\bar{l}\gamma_\mu P_L \chi^{(n)} + h.c. \Big)\,, \nonumber \\
\mathcal{L}_{G^\pm}^{int} &= -\dfrac{g_w}{\sqrt{2} M_W} G^-  \sum_{l=e, \mu, \tau}
\Big[B_{l\nu_l}m_l\bar{l} P_L \nu_l + \sum_{n=-\infty}^{\infty}  B_{l, n}\bar{l}
\big(m_l P_L - m_{(n)}P_R\big) \chi^{(n)} + h.c.\Big]\,,
\end{align}  
where $g_w$ is the weak coupling constant and $P_{R, L}=\dfrac{1\pm\gamma_5}{2}$ are the chirality projection operators. Here $m_{(n)}$ and $m_l$ represent the masses of KK neutrinos and charged leptons respectively. Also the expressions for the elements of the matrix $B$ are given in~\cite{Ioannisian:1999cw}
\begin{align}
\label{eq: VU}
 B_{l,n}=\sum_{i=e, \mu, \tau} V_{li}^{l} U_{i,n}^{\nu}\,,\qquad B_{l\nu_k}= \sum_{i=e, \mu, \tau} V_{li}^{l} U_{ik}^{\nu}  \quad \textrm{for each}\:\: k=e, \mu, \tau\,,  
\end{align}
where the matrix $V^l$ diagonalizes the charged lepton mass matrix
 Indeed, the KK neutrino mixing parameters emphasized in~\cite{Ioannisian:1999cw,Langacker,jcaban} are
\begin{align}
\label{eq:sinemixing}
(s_{L}^{\nu_i})^2=\sum_{n=1}^{+\infty} {|B_{l,n}|^{2}} \approx \Xi^{(i)} \Xi^{(i)T}\approx \sum_{n=1}^{\infty} { U_{0n}^{(i)2}}.
\end{align}
The discrete summation including all of the KK modes can be written into continuous integration over all of its energy scale $E$ as a prescription of~\cite{Ioannisian:1999cw,jcaban} 
\begin{align}
\label{eq:discont}
\sum_{n=1}^{\infty} \longrightarrow S_\delta R^\delta \int_{\frac{1}{R}}^{M_F} E^{\delta-1} dE\,,
\end{align}
with $M_F$ as the ultraviolet (UV) cut-off, $R$ is the radius of the extra dimension, and $S_\delta=2\pi^{\delta/2}/\Gamma(\frac{\delta}{2})$ to be the surface area of the unit sphere in $\delta$ dimensions. As a result, the mixings can be expressed as
\begin{align}
\label{eq:sinemix2}
(s_{L}^{\nu_i})^2\approx  \begin{cases} 
      \dfrac{\pi h_{i}^{2}v^2 }{ M_{F}^{2} } \ln \Big[ \dfrac{ M_{P}^{2}}{ M_{F}^{2} }\Big]  
& \textrm{for}\:\:\: \delta=2\\
\dfrac{S_\delta h_{i}^{2}v^2 }{ (\delta-2)M_{F}^{2}} \Big[\Big(1 - \big(\dfrac{M_F}{M_P}\big)^{2 - \frac{4}{\delta}} \Big ) \Big]  & \textrm{for}\:\:\: \delta>2\,. 
   \end{cases}
\end{align}

\section{$b\rightarrow c \tau \nu_{\tau}^{KK}$ transition and constraints}
 \label{sec:btoctaunu}
\begin{figure}[t]
	\centering
	\includegraphics[width=0.5\textwidth]{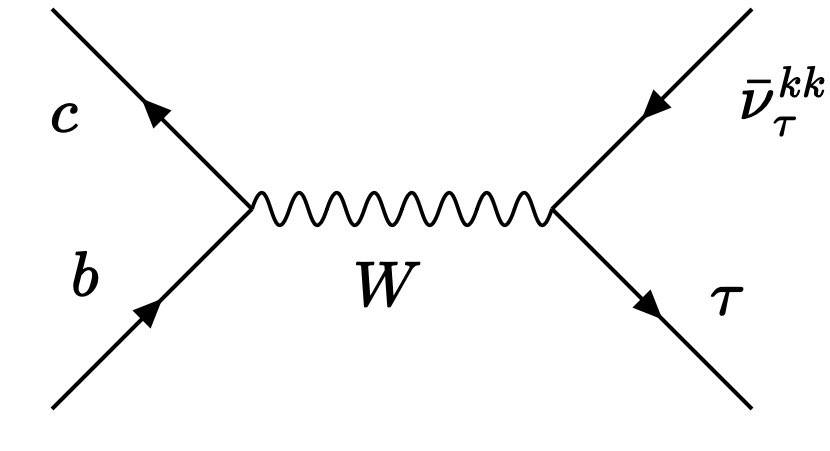}
	\caption{The relevant Feynman diagram of $b \rightarrow c \tau \bar{\nu}_{\tau}^{KK}$ transition that contributes to the same experimental signatures of $B\to D^{(*)}\tau\bar{\nu}$ decay.}
	\label{fig:bstaukk}
\end{figure}

\begin{figure}[t]
	\centering
	\includegraphics[width=1\textwidth]{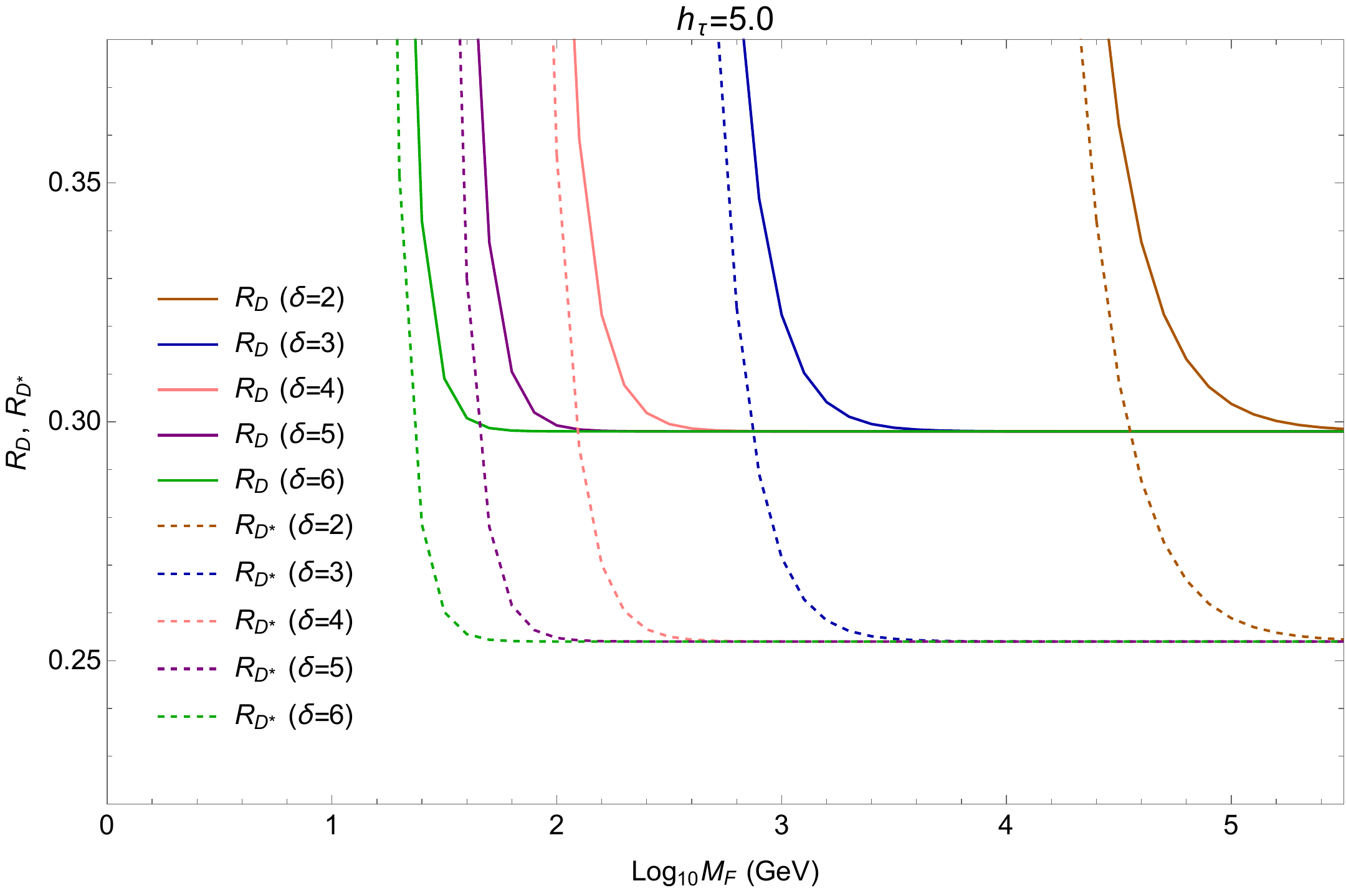}
	\caption{The plot of $M_F$ vs $R_{D^{(*)}}$ for a fixed Yukawa coupling strength $h_\tau=5$. The solid lines represent the $R_D$ while the dashed lines represent the $R_{D^*}$. }
	\label{fig:mfvsrdrdstar}
\end{figure}

In additional to the SM diagram, the relevant Feynman diagram in extra-dimensional model that gives the the same experimental signatures as SM $B\to D^{(*)}\tau \bar{\nu}$ is  shown in Fig. \ref{fig:bstaukk}.
Therefore, together with the new contributions from the KK neutrinos, the prediction for $R_{D^{(*)}}$ is 
\begin{align}
\label{eq:rd}
R_{D^{(*)}}&=\dfrac{Br(B\rightarrow D^{(*)}\tau \bar{\nu}_{\tau} )+Br(B\rightarrow D^{(*)}\tau \bar{\nu}_{\tau}^{KK})}{[(Br(B\rightarrow D^{(*)}e \bar{\nu}_{e} + Br(B\rightarrow D^{(*)}\mu \bar{\nu}_{\mu} ))/2]} \approx R_{D^{(*)}}^{SM}\left( 1+ \frac{\Gamma(B\rightarrow D^{(*)}\tau \bar{\nu}_{\tau}^{KK}))}{\Gamma^{SM}(B\rightarrow D^{(*)}\tau \bar{\nu}_{\tau}))}\right)\\\nonumber
&\approx R_{D^{(*)}}^{SM}\left( 1+ \ \sum_{n=1}^{+\infty} \eta_n B_{\tau,n}^{*} B_{\tau,n}\right) =R_{D^{(*)}}^{SM}\left( 1+ \dfrac{h_{\tau}^{2} v^2 M_{P}^{\frac{4}{\delta}-2}}{M_{F}^{\frac{4}{\delta}}} S_\delta R^{\delta-2}\int_{\frac{1}{R}}^{m_{B}-m_{D^{(*)}}-m_{\tau}} \frac{E^{\delta-1}}{m^2 + E^2} \eta(E) dE\right), 
\end{align}
where $m^2=\dfrac{h_{\tau}^{2} v^2 M_{F}^{2}}{2 M_{P}^{2}}$ and $\eta_n$ are the effects from three-body phase, having $\eta(E)$ as the corresponding transformation of $\eta_n$ (see the Appendix for the details) after summing up over all the contributions from KK neutrinos while invoking Eq.~\eqref{eq:discont}.
Also note that we have $\dfrac{1}{R}\simeq \dfrac{ M_{F}^{\frac{2}{\delta}+1}}{M_{P}^{\frac{2}{\delta}}}$ from the fundamental relation between $M_{P}$ and $M_{F}$. To fit the $R_D$ central value, the needed $M_F$ is about $7~{\rm TeV}$ for Yukawa coupling $h_\tau=1$, which is excluded by the LHC mono-jet plus missing energy search~\cite{ATLAS:2021kxv} that imposes a lower bound $M_F \gtrsim 11.2~{\rm TeV}$.
Therefore, we choose Yukawa coupling $h_\tau=5$ for our benchmark value throughout this paper unless otherwise stated. Fig.~\ref{fig:mfvsrdrdstar} shows the relation between fundamental scale $M_F$ and $R_{D^{(*)}}$.
The plots are made by plugging in the SM central values of $R_{D^{(*)}}^{SM}$~\cite{HFLAV:2022pwe} in eq.~\eqref{eq:rd}. The solid and dashed lines represent the $R_{D}$ and $R_{D^*}$, respectively, for $\delta=2,3,4,5,$ and 6. Clearly as fundamental scale $M_F$ increases, $R_{D^{(*)}}$ approach to the SM values. The best fit of the most recent experimental central values of $R_{D^{(*)}}^{exp}$~\cite{HFLAV:2023prelim} with Yukawa coupling $h_\tau=5$ are
\begin{align}
\label{eq:MFvaluesDeltaRD}
M_{F}=33\: \textrm{TeV},\: 748 \: \textrm{GeV},\: 127 \: \textrm{GeV}, \:46\: \textrm{GeV},\: 24\: \textrm{GeV} \:\:\:\:\:\textrm{for}\:\:R_{D}\,, 
\end{align}
\begin{align}
\label{eq:MFvaluesDeltaRDstar}
M_{F}=42\: \textrm{TeV},\: 834 \: \textrm{GeV},\: 135\: \textrm{GeV}, \:48\: \textrm{GeV},\: 24\: \textrm{GeV} \:\:\:\:\:\textrm{for}\:\: R_{D^*}\,,
\end{align}
corresponding to $\delta=2,3,4,5,6$. 
It is clear that the fundamental scales $M_F$ are excluded by LHC searches, even when $h_\tau$ is increased up to $4\pi$, except for $\delta=2$. 
Hence, the only feasible scenario is when $\delta=2$. 
\begin{figure}[t]
	\centering
	\includegraphics[width=0.7\textwidth]{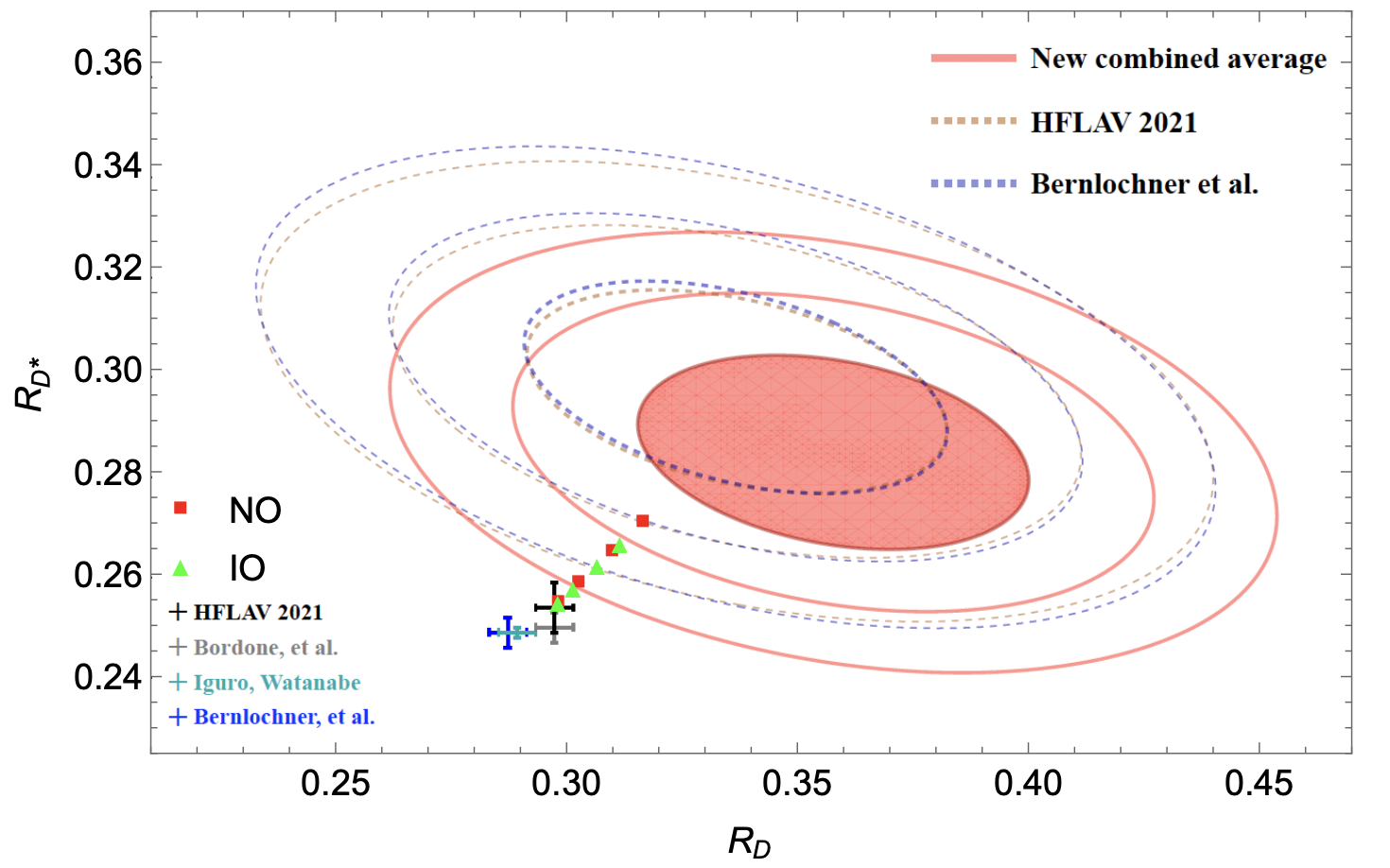}
	\includegraphics[width=0.7\textwidth]{rdvsrdstarv2}	
	\caption{The upper panel shows the ($1\sigma, 2\sigma, 3\sigma$) contour plots of the experimental results of $R_{D}$ and $R_{D^*}$ of the world average (red)~\cite{Iguro} in mid-autumn 2022, \cite{HFLAVCollaboration} HFLAV 2021 average results (dashed orange) and with~\cite{Bernlochner}(dashed purple). The Standard Model (SM) predictions are denoted by the crosses~\cite{UBernlochner,Watanabe,Bordonea,Bordoneb}. The pairs of points $(R_{D}, R_{D^*})$ in normal (inverted) ordering are shown in red squares (green triangles) with Yukawa couplings $h_\tau=1,5,8,10$ and fixed $M_{F}=110 (128)$ TeV. The lower panel shows the most updated results  of the combined new world average using the most recent results of LHCb~\cite{HFLAV:2023prelim}, and the global picture is almost the same as in upper panel.}
	\label{fig:hflav}
\end{figure} 

The Fig.~\ref{fig:hflav} shows the ($1\sigma, 2\sigma, 3\sigma$) contour plots between $R_D$ and $R_{D^*}$ taken from~\cite{Iguro,HFLAVCollaboration,Bernlochner}. The Standard Model predictions are the crosses~\cite{UBernlochner,Watanabe,Bordonea,Bordoneb}. Along with the $KK$ neutrinos contributions to $R_{D}$ and $R_{D^*}$ the four pairs of points ($R_{D}, R_{D^*}$) are plotted corresponding to Yukawa couplings $h_\tau=1,5,8,10$ with fixed   $M_F=110$ TeV and $M_F=128$ TeV derived from the lower bounds~\cite{Forero:2022skg} in neutrino fittings for normal and inverted ordering, respectively. 
It can be observed that as Yukawa coupling increases the contributions from KK neutrinos increase, and as a result, predictions of $R_{D^{(*)}}$ are more close to the experimental central values of the new world average~\cite{Iguro}. 
 
In the case of normal ordering the two points corresponding to $h_\tau=5$ and $h_\tau=8$  lie within the $3\sigma$ of the world average, while the point with $h_\tau=1$ lies outside $3\sigma$, and lastly the point with $h_\tau=10$ lies within $2\sigma$. Moreover, in inverted ordering the points with $h_\tau=8$ and $h_\tau=10$ lie within $3\sigma$, and the remaining two points lie outside $3\sigma$. With regards to the most updated results from LHCb shown in lower panel, the global picture of the combined world average will not change.

\section{Constraints from $\tau$ decays}
\label{sec:constraints}
Now let's consider constraints for the predictions of $R_{D}$ and $R_{D^*}$. Since the new contributions come from the $\tau$ lepton sector only, the most relevant and stringent constraints we consider are the experimental bounds for rare $\tau$ decays, including $\tau\rightarrow e\gamma$, $\tau\rightarrow \mu\gamma$,  $\tau\rightarrow e\mu^+\mu^-$ and $\tau\rightarrow \mu e^+e^-$.

After summing up all the KK neutrino modes, the expression of the branching ratios of $\tau \rightarrow l^{'}\gamma$ is given as~\cite{Ioannisian:1999cw}
\begin{align}
\label{eq:Bllpg}
Br(l\rightarrow l^{'} \gamma)  \approx  \frac{\alpha^{3}_{w}\, s^{2}_{w}}{1024\pi^{2}} \, \frac{m^{4}_{\tau}}{M^{4}_{W}}\, \frac{m_{\tau}}{\Gamma_{\tau}} \, (s^{\nu_{\tau}}_{L})^{2}\, (s^{\nu_{l^{'}}}_{L})^{2}\,,
\end{align}
where  $M_W$ is the mass of $W$-boson, $\alpha_w=\frac{g_w^2}{4\pi}$ with $g_w$ being the $SU(2)_L$ coupling strength, $s_w=\sin\theta_w$ with $\theta_w$ being the weak mixing angle, and $\ell^\prime=e,~\mu$.
The current experimental value from, $Br_{\text{exp}}(\tau \rightarrow e \gamma) < 3.3 \times 10^{-8}$ and $Br_{\text{exp}}(\tau \rightarrow \mu \gamma) < 4.2 \times 10^{-8}$ \cite{Workman:2022ynf} at 90$\%$ confidence level (CL), and $\Gamma_\tau=2.23 \times 10^{-12}$ GeV using the mean-lifetime of $\tau$ lepton in~\cite{Workman:2022ynf}, one obtains the lower limits 
\begin{align}
M_F >  \begin{cases} 
      67  \:\textrm{TeV}
& \textrm{for}\:\:\: \tau \rightarrow e \gamma\,, \\
63   \:\textrm{TeV} & \textrm{for}\:\:\: \tau \rightarrow \mu \gamma\,, 
   \end{cases}
\end{align}
by setting $ h_e=h_\mu=1$ and  $h_\tau=5$.
The next constraints come from the three-body decay of $\tau$ lepton, namely $\tau\rightarrow e\mu\mu$ and $\tau\rightarrow \mu ee$. 
The corrresponding branching ratios are given by~\cite{Ioannisian:1999cw}
\begin{align}
\label{eq:Btauemumu}
Br(\tau \rightarrow e \mu\mu)& =  \dfrac{\alpha_{w}^{4}}{98304} \dfrac{m_{\tau}^{4}}{M_{W}^{4}} \dfrac{M_{F}^{4}}{M_{W}^{4}} d_{\delta}^{2} (s_{L}^{\nu_\tau})^2(s_{L}^{\nu_e})^2 \Bigg\{(s_{L}^{\nu_\mu})^4
+ 2(1-2s_{w}^{2})(s_{L}^{\nu_\mu})^4\big[\sum_{l=e,\mu,\tau}{(s_{L}^{\nu_l})^2} \big]\nonumber \\
&+ 8s_{w}^{4 }\big[ \sum_{l=e,\mu,\tau}(s_{L}^{\nu_l})^2\big]^2
 \Bigg\}\,,
\end{align}
\begin{align}
\label{eq:Btaumumue}
Br(\tau \rightarrow \mu e e)& =  \dfrac{\alpha_{w}^{4}}{98304} \dfrac{m_{\tau}^{4}}{M_{W}^{4}} \dfrac{M_{F}^{4}}{M_{W}^{4}} d_{\delta}^{2} (s_{L}^{\nu_\tau})^2(s_{L}^{\nu_\mu})^2 \Bigg\{(s_{L}^{\nu_e})^4
+ 2(1-2s_{w}^{2})(s_{L}^{\nu_e})^4\big[\sum_{l=e,\mu,\tau}{(s_{L}^{\nu_l})^2} \big]\nonumber \\
&+ 8s_{w}^{4 }\big[ \sum_{l=e,\mu,\tau}(s_{L}^{\nu_l})^2\big]^2
 \Bigg\}\,,
\end{align}
where $d_\delta = d_2=\frac{\pi^2}{12\ln^2(M_P/M_F)}\lesssim 1$ is  a dimension-dependent factor. The current experimental bounds $Br(\tau \rightarrow e \mu\mu)<2.7\times 10^{-8} $  and $Br(\tau \rightarrow \mu ee)<1.8\times 10^{-8}$~\cite{Workman:2022ynf} give 
\begin{align}
\label{eq:Btaumumue}
M_F >  \begin{cases} 
      54  \:\textrm{TeV}
& \textrm{for}\:\:\: \tau \rightarrow e \mu\mu\,, \\
60   \:\textrm{TeV} & \textrm{for}\:\:\: \tau \rightarrow \mu ee\,, 
   \end{cases}
\end{align}
with $h_\tau=5$ and  $h_e=h_\mu=1$.
The last constraints are from the fittings of various neutrino experiments.  It is shown that the upper bounds on the size of the extra dimension should be smaller than $0.2~{\rm \mu m}$ and $0.1~{\rm \mu m}$ at $90~\%$ C.L. for normal and inverted ordering, respectively~\cite{Forero:2022skg}.

The corresponding lower limits of fundamental scale for $\delta =2$ are given as 
\begin{align}
\label{eq:neutrinoboundslimits}
M_{F} > 110\:(128)\: \textrm{TeV}\,,
\end{align}
for normal ordering (inverted ordering).
\begin{figure}[t]
	\centering
	\includegraphics[width=0.45\textwidth]{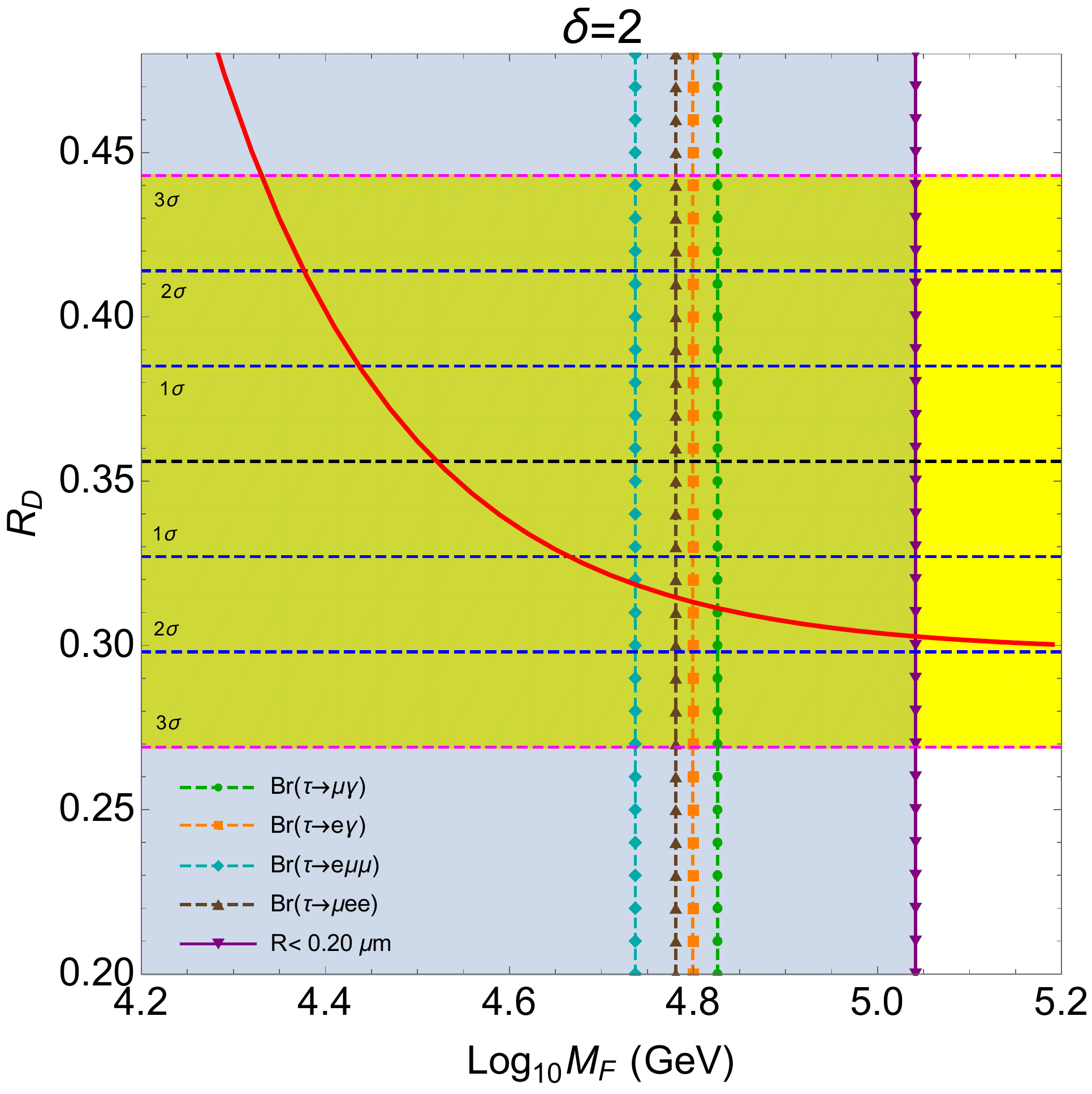}\hspace*{0.1cm}
	\includegraphics[width=0.45\textwidth]{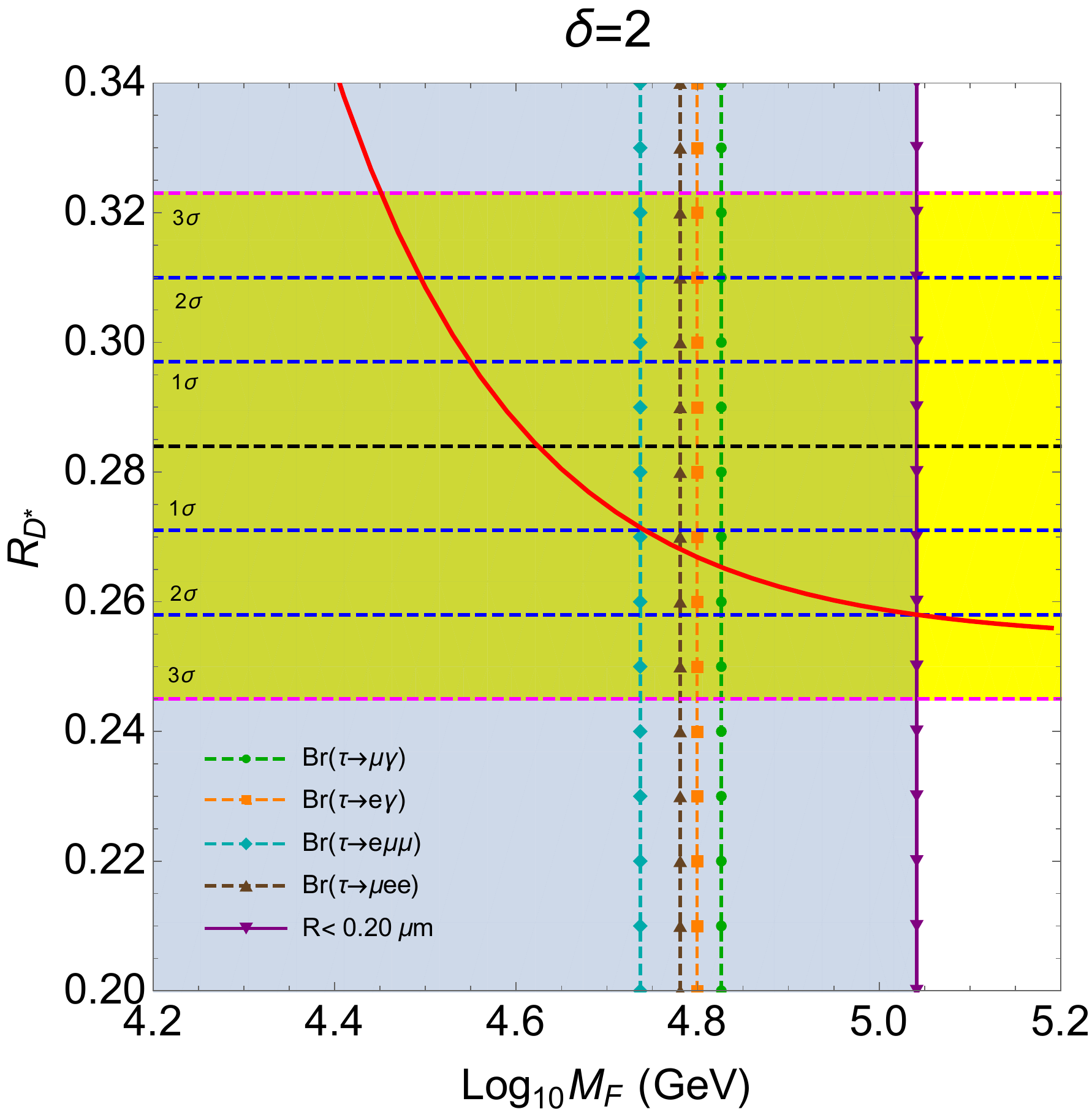}
	\includegraphics[width=0.15\textwidth]{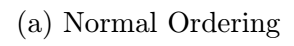}
	\vspace*{0.05cm}
	\includegraphics[width=0.45\textwidth]{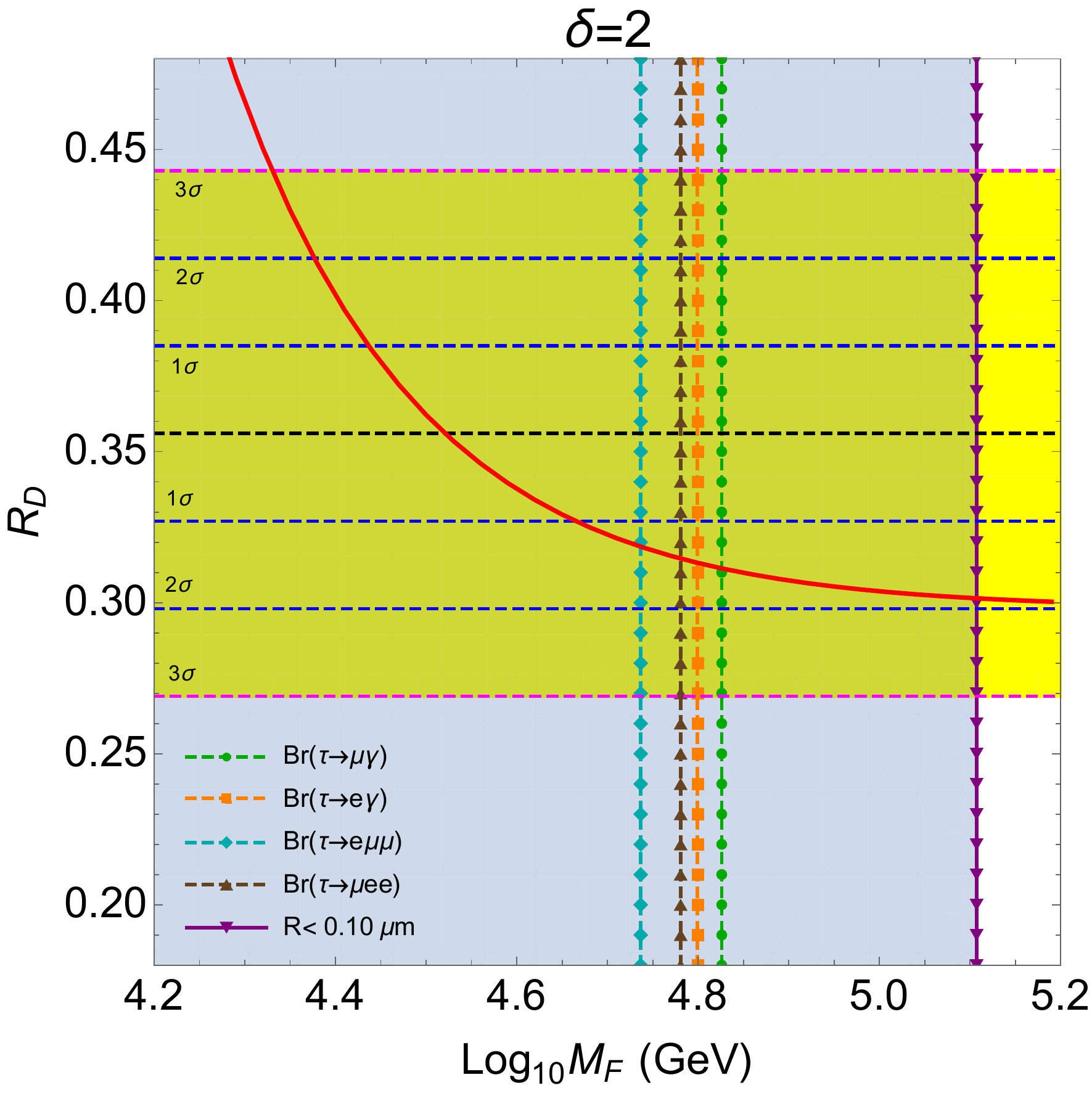}\hspace*{0.1cm}
	\includegraphics[width=0.45\textwidth]{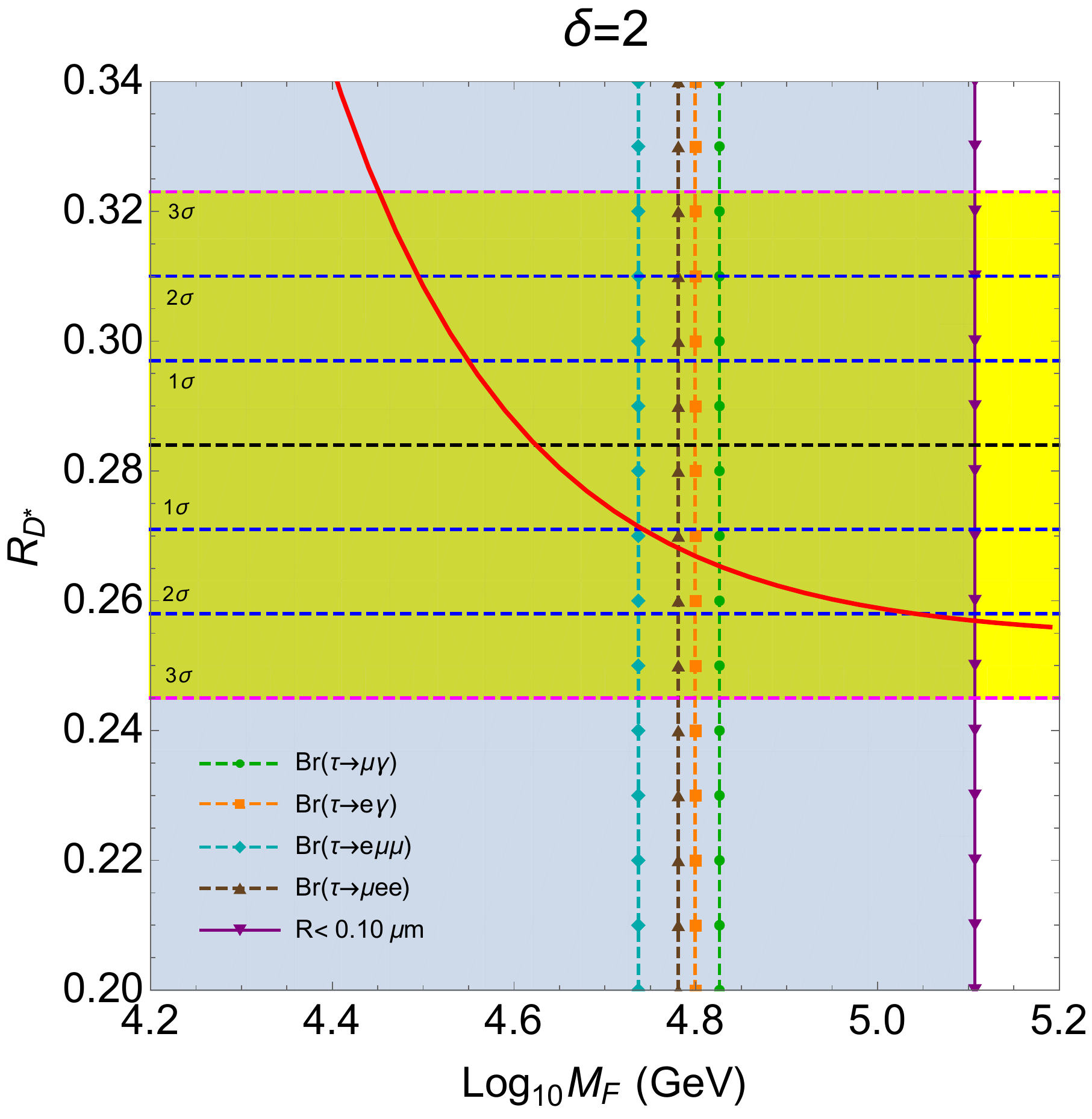} 
	\includegraphics[width=0.15\textwidth]{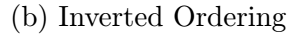}
\caption{The left (right) panel of Fig.~\ref{fig:rdvsrdstar}(a) and Fig.~\ref{fig:rdvsrdstar}(b)  is the plot of $M_F$ vs $ R_{D^{(*)}}$  for $\delta=2$ with Yukawa coupling $h_\tau=5$ together with the following constraints:                                                                                                                  (i.) Br$(\tau\rightarrow e\mu\mu)<2.7\times 10^{-8}$ (dashed cyan), (ii.) Br$(\tau\rightarrow \mu ee)<1.8\times 10^{-8}$ (dashed brown) , (iii.) Br$(\tau\rightarrow \mu\gamma)<4.2\times 10^{-8}$ (dashed green), 
(iv.) Br$(\tau\rightarrow e\gamma)< 3.3\times 10^{-8}$ (dashed orange)~\cite{Workman:2022ynf}, and (v.) neutrino bounds (solid purple)~\cite{Forero:2022skg}. The yellow bands give $1\sigma$, $2\sigma$ and $3\sigma$ regions of $R_{D^{(*)}}^{exp}$. The dashed black lines determine the central values of $R_{D^{(*)}}^{exp}$, while dashed blue and red lines are the boundaries of $1\sigma$ and $2\sigma$ , and $3\sigma$ regions of $R_{D^{(*)}}^{exp}$. The data we used here is the most updated world average~\cite{HFLAV:2023prelim}. }
\label{fig:rdvsrdstar}
\end{figure} 
Fig.~\ref{fig:rdvsrdstar} summarises the data of $R_{D^{(*)}}$ for $\delta=2$ with Yukawa coupling $h_\tau=5$ together with the experimental bounds on $M_F$ from
$Br(\tau\rightarrow e\mu\mu)$, $Br(\tau\rightarrow \mu ee)$,  
$Br(\tau\rightarrow \mu\gamma)$, $Br(\tau\rightarrow e\gamma)$ and neutrino oscillations.
The yellow bands determine the $1\sigma$, $2\sigma$  and $3\sigma$ regions of $R_{D^{(*)}}$. The horizontal dashed lines give the central values, and boundaries of $1\sigma$, $2\sigma$ and  $3\sigma$ of $R_{D^{(*)}}^{exp}$. The most stringent bound comes from the size of extra dimensions for normal and inverted ordering  in~\cite{Forero:2022skg}. As determined in~\eqref{eq:neutrinoboundslimits}, these lower limits correspond to  $R_{D}=0.304\:(0.301)$ and $R_{D^*}=0.259\:(0.257)$, for  normal (inverted) ordering, respectively. All these $R_{D^{(*)}}$ predictions can be found very near to the boundary of $2\sigma$ below from the central values of $R_{D^{(*)}}^{exp}$ for both NO and IO, respectively.
\section{Conclusion }
\label{sec:con}
A possible violation of the lepton flavor universality can be found in anomalies involving rare B meson decays. This is a positive sign of physics beyond the SM. A newly calculated world average of the data by different experimental groups BaBar, Belle, and LHCb collaboration strongly supports again the leptonic flavor universality violation in the $b\rightarrow c\tau\bar{\nu}_\tau$ transition. We show in this work that it is possible to explain these anomalies in the extra-dimensional framework, where the Planck scale $M_P$ is lowered to the fundamental scale $M_F$. By introducing three right-handed neutrinos propagating in the bulk, the contributions from their corresponding KK neutrino modes after compactification give a plausible description of the anomalies through mixings from the active neutrinos. The central values $R_{D}^{exp}=0.356$ and $R_{D^*}^{exp}=0.284$ ruled out the cases $\delta=3,4,5,$ and 6 since the needed values of $M_F$ are lower than the bounds from LHC searches.  As a result, we only considered the very special number of extra dimensions $\delta=2$. The most severe bounds from neutrino experiments on the size of large extra dimension are $R<0.2\mu m$ and $R<0.1 \mu m$ for NO and IO, respectively. To satisfy these bounds the lower limits for the fundamental scale $M_F$ must be 110 TeV and 128 TeV, for NO and IO, respectively. With Yukawa coupling strength $h_\tau=5$, the predictions for $R_{D}$ and $R_{D^*}$ with the corresponding lower limits of $M_F$ from neutrino experiments  are 0.304 (0.301) and 0.259 (0.257) on the boundary of $2\sigma$ contour, respectively, for NO (IO). Apparently, there is a tension between central values of $R_{D^{(*)}}^{exp}$ with the lower bounds from neutrino experiments. The future measurements of $R_{D^{(*)}}^{exp}$  will exclude this extra dimensional model with right-handed neutrino propagating in the bulk, if the central values stay.

\section*{Acknowledgment}  
We would like to acknowledge the support of National Center for Theoretical Sciences (NCTS). This work was supported in part by the National Science and Technology Council (NSTC) of Taiwan under Grant No.MOST 110-2112-M-003-003-, 111-2112-M-003-006 and 111-2811-M-003-025-.


\section*{Appendix}
\label{appendix}  
The three-body phase of $B \to D \tau \nu_\tau^{(n)KK}$ is given as 

\begin{align}
\label{eq:decaywidth}
\int_{y^{(n)}_{min}}^{y_{max}} \int_{x^{(n)}_{min}}^{x^{(n)}_{max}} (x+y+s)(t^{(n)}-x-y) dxdy\,,
\end{align} 
where
\begin{align}
\label{eq:form1}
s=-(m_{D^{(*)}}^{2}+m_{\tau}^{2}), \:\:\:\:\ t^{(n)}=(m_{B}^{2} + m_{KK}^{(n)2}) \,.
\end{align}.
The lower and upper limits in variable $y$ are
\begin{align}
\label{eq:ylimits}
y_{min}^{(n)}=(m^{(n)}_{KK}+m_\tau)^2, \:\:\:\:\  y_{max}=(m_{B} -m_{D^{(*)}})^2\,,
\end{align}
while the lower and upper limits for variable $x$ can be written in terms of $y$
\begin{align}
\label{eq:xlimits}
x^{(n)}_{max/min}=\dfrac{-(y^2+A^{(n)}-yB^{(n)})\pm \sqrt{(y^2+A^{(n)}-yB^{(n)})^2 -4y(yC^{(n)}+D^{(n)})}}{2y}\,,
\end{align}
such that the expressions for $A^{(n)}, B^{(n)}, C^{(n)}$, and $D^{(n)}$ are the following:
\begin{align}
\label{eq:form2}
A^{(n)}&=(m_{\tau}^{2}-m_{KK}^{(n)2})(m_{B}^{2}-m_{D^{(*)}}^{2})\,,\\
B^{(n)}&=(m_{B}^{2}+ m_{D^{(*)}}^{2}+m_{KK}^{(n)2}+ m_{\tau}^{2})\,,\\
C^{(n)}&=(m_{D^{(*)}}^{2}-m_{KK}^{(n)2})(m_{B}^{2}-m_{\tau}^{2})\,,\\
D^{(n)}&=(m_{KK}^{(n)2} m_{B}^{2} - m_{D^{(*)}}^{2}m_{\tau}^{2})(m_{B}^{2}-m_{D^{(*)}}^{2}+m_{KK}^{(n)2}- m_{\tau}^{2}).
\end{align}
Here $m_B$, $m_{D^{(*)}}$, $m^{(n)}_{KK}$, and $m_{\tau}$ are the masses of the B-meson, D-meson, KK mass eigenstates, and tau lepton respectively. 
When we sum over $n$, the integration replacement of the discrete sum in Eq.~\eqref{eq:discont} transforms Eq.~\eqref{eq:decaywidth} into 
\begin{align}
\label{eq:decaywidthnewform}
S_\delta R^{\delta-2}\int_{\frac{1}{R}}^{m_{B}-m_{D^{(*)}}-m_{\tau}} \frac{E^{\delta-1}}{m^2 + E^2}
\times \int_{y_{min}}^{y_{max}} \int_{x_{min}}^{x_{max}} (x+y+s)(t-x-y) dx dy dE\,,
\end{align} 
such that $m^2=\dfrac{h_{\tau}^{2} v^2 M_{F}^{2}}{2 M_{P}^{2}}$ and every appearance of $m^{(n)}_{KK}$ in the expression  
\begin{align}
\label{eq:replacekk}
\int_{y^{(n)}_{min}}^{y_{max}} \int_{x^{(n)}_{min}}^{x^{(n)}_{max}} (x+y+s)(t^{(n)}-x-y) dx dy\,,
\end{align} 
is replaced by variable $E$.
The prediction
for $R_{D^{(*)}}$ with contributions from the KK neutrinos is given by 
\begin{align}
\label{eq:rd}
R_{D^{(*)}}&=\dfrac{Br(B\rightarrow D^{(*)}\tau \bar{\nu}_{\tau} )+Br(B\rightarrow D^{(*)}\tau \bar{\nu}_{\tau}^{KK})}{[(Br(B\rightarrow D^{(*)}e \bar{\nu}_{e} + Br(B\rightarrow D^{(*)}\mu \bar{\nu}_{\mu} ))/2]} \approx R_{D^{(*)}}^{SM}\left( 1+ \frac{\Gamma(B\rightarrow D^{(*)}\tau \bar{\nu}_{\tau}^{KK}))}{\Gamma^{SM}(B\rightarrow D^{(*)}\tau \bar{\nu}_{\tau}))}\right)\\\nonumber
&\approx R_{D^{(*)}}^{SM}\left( 1+ \ \sum_{n=1}^{+\infty} \eta_n B_{\tau,n}^{*} B_{\tau,n}\right) =R_{D^{(*)}}^{SM}\left( 1+ \dfrac{h_{\tau}^{2} v^2 M_{P}^{\frac{4}{\delta}-2}}{M_{F}^{\frac{4}{\delta}}} S_\delta R^{\delta-2}\int_{\frac{1}{R}}^{m_{B}-m_{D^{(*)}}-m_{\tau}} \frac{E^{\delta-1}}{m^2 + E^2} \eta(E) dE\right), 
\end{align}
where 
\begin{align}
\label{eq:solrd}
\eta_n =\dfrac{\int_{y^{(n)}_{min}}^{y_{max}} \int_{x^{(n)}_{min}}^{x^{(n)}_{max}} (x+y+s)(t^{(n)}-x-y) dx dy}{\int_{y_{min}}^{y_{max}} \int_{x_{min}}^{x_{max}} (x+y+s)(t-x-y) dx dy \:\:\Big|_{SM}},~~
\eta(E)=\eta_n\:\:\Big|_{m^{(n)}_{KK}=E}\,.
\end{align}


\begin{thebibliography}{99}



\bibitem{LHCb:2022qnv}
 [LHCb],
[arXiv:2212.09152 [hep-ex]].


\bibitem{HFLAV:2022pwe}
Y.~Amhis \textit{et al.} [HFLAV],
[arXiv:2206.07501 [hep-ex]].

\bibitem{Iguro}
S. Iguro, T. Kitahara and R. Watanabe, doi:10.48550/arXiv.2210.10751
[arXiv:2210.10751 [hep-ph]]




\bibitem{HFLAV:2023prelim}
https://indico.cern.ch/event/1231797/




\bibitem{Hiller:2021pul}
G.~Hiller, D.~Loose and I.~Ni\v{s}and\v{z}i\'c,
JHEP \textbf{06}, 080 (2021);
B.~Gripaios, M.~Nardecchia and S.~A.~Renner,
JHEP \textbf{05}, 006 (2015);
R.~Barbieri, C.~W.~Murphy and F.~Senia,
Eur. Phys. J. C \textbf{77}, no.1, 8 (2017);
B.~Fornal, S.~A.~Gadam and B.~Grinstein,
Phys. Rev. D \textbf{99}, no.5, 055025 (2019);
C.~Cornella, J.~Fuentes-Martin and G.~Isidori,
JHEP \textbf{07}, 168 (2019);
O.~Popov, M.~A.~Schmidt and G.~White,
Phys. Rev. D \textbf{100}, no.3, 035028 (2019);
%
I.~Bigaran, J.~Gargalionis and R.~R.~Volkas,
JHEP \textbf{10}, 106 (2019);
%
%
C.~Hati, J.~Kriewald, J.~Orloff and A.~M.~Teixeira,
JHEP \textbf{12}, 006 (2019);
A.~Datta, J.~L.~Feng, S.~Kamali and J.~Kumar,
Phys. Rev. D \textbf{101}, no.3, 035010 (2020);
P.~S.~Bhupal Dev, R.~Mohanta, S.~Patra and S.~Sahoo,
Phys. Rev. D \textbf{102}, no.9, 095012 (2020);
M.~Du, J.~Liang, Z.~Liu and V.~Q.~Tran,
K.~Ban, Y.~Jho, Y.~Kwon, S.~C.~Park, S.~Park and P.~Y.~Tseng,








\bibitem{Megias:2017ove}
E.~Megias, M.~Quiros and L.~Salas,
JHEP \textbf{07}, 102 (2017);
X.~G.~He and G.~Valencia,
Phys. Lett. B \textbf{779}, 52-57 (2018);
S.~Matsuzaki, K.~Nishiwaki and R.~Watanabe,
JHEP \textbf{08}, 145 (2017);
K.~S.~Babu, B.~Dutta and R.~N.~Mohapatra,
JHEP \textbf{01}, 168 (2019);
A.~Greljo, D.~J.~Robinson, B.~Shakya and J.~Zupan,
JHEP \textbf{09}, 169 (2018);
P.~Asadi, M.~R.~Buckley and D.~Shih,
JHEP \textbf{09}, 010 (2018)


\bibitem{Tanaka:1994ay}
M.~Tanaka,
Z. Phys. C \textbf{67}, 321-326 (1995);
A.~Celis, M.~Jung, X.~Q.~Li and A.~Pich,
JHEP \textbf{01}, 054 (2013);
A.~Celis, M.~Jung, X.~Q.~Li and A.~Pich,
Phys. Lett. B \textbf{771}, 168-179 (2017);
S.~Iguro and K.~Tobe,
Nucl. Phys. B \textbf{925}, 560-606 (2017);
S.~Fraser, C.~Marzo, L.~Marzola, M.~Raidal and C.~Spethmann,
Phys. Rev. D \textbf{98}, no.3, 035016 (2018);
R.~Martinez, C.~F.~Sierra and G.~Valencia,
Phys. Rev. D \textbf{98}, no.11, 115012 (2018);


\bibitem{Strumia:2000}
R.~Barbieri, P.~Creminelli and A.~Strumia
Nucl. Phys. B \textbf{585}, 28-44 (2000)
doi:10.1016/S0550-3213(00)00348-5
[arXiv:hep-ph/0002199 [hep-ph]]


\bibitem{Forero:2022skg}
D.~V.~Forero, C.~Giunti, C.~A.~Ternes and O.~Tyagi,
Phys. Rev. D \textbf{106}, no.3, 035027 (2022)
doi:10.1103/PhysRevD.106.035027
[arXiv:2207.02790 [hep-ph]].


\bibitem{Hamed} 
N.~Arkani-Hamed, S.~Dimopoulos, G. Dvali and J. March-Russell,
Phys. Rev. D \textbf{65}, 024032 (2001) 
doi:10.1103/PhysRevD.65.024032 
[arXiv:hep-ph/9811448 [hep-ph]].

\bibitem{Dienes} 
K.R. Dienes, E. Dudas and T. Gherghetta, 
Nucl. Phys. B \textbf{557}, 25 (1999)
doi:10.1016/S0550-3213(99)00377-6
[arXiv:hep-ph/9811428 [hep-ph]].
 
\bibitem{Dvali}
G. R. Dvali and A. Y. Smirnov, Nucl. Phys. B \textbf{563}, 63 (1999) 
doi:10.1016/S0550-3213(99)00574-X
[arXiv:hep-ph/9904211 [hep-ph]]



\bibitem{Ioannisian:1999cw}
A.~Ioannisian and A.~Pilaftsis,
Phys. Rev. D \textbf{62}, 066001 (2000)
doi:10.1103/PhysRevD.62.066001
[arXiv:hep-ph/9907522 [hep-ph]].


 
 



\bibitem{Mohapatra}
R. N. Mohapatra and A. Perez-Lorenzana, Nucl. Phys. B \textbf{593}, 451 (2001)
doi:10.1016/S0550-3213(00)00634-9
[arXiv:hep-ph/0006278  [hep-ph]] 

\bibitem{Langacker}
P. Langacker, D. London, Phys. Rev. D \textbf{38}, 886 (1988)  
doi:org/10.1103/PhysRevD.38.886.


\bibitem{jcaban}
J.C. Aban, C.R. Chen and C.S. Nugroho, Phys. Lett. B \textbf{830}, 137164 (2022)
doi:10.1016/j.physletb.2022.137164 
[arXiv:hep-ph/2112.12477 [hep-ph]]


\bibitem{Schechter}
J. Schechter and J.W.F. Valle, Phys. Rev. D \textbf{22}, 2227 (1980)  
doi:10.1103/PhysRevD.22.2227.


\bibitem{ATLAS:2021kxv}
G.~Aad \textit{et al.} [ATLAS],
Phys. Rev. D \textbf{103}, no.11, 112006 (2021)
doi:10.1103/PhysRevD.103.112006
[arXiv:2102.10874 [hep-ex]].





\bibitem{HFLAVCollaboration}
HFLAV Collaboration,
doi:10.48550/arXiv.2206.07501 
[arXiv:2206.07501 [hep-ex]]

\bibitem{Bernlochner}
F. U. Bernlochner, M. F. Sevilla, D. J. Robinson, and G. Wormser, 
Rev. Mod. Phys. \textbf{94}, 015003 (2022) 
[arXiv:2101.08326] [hep-ex]]


\bibitem{UBernlochner}
F.~U.~Bernlochner, M.~F.~Sevilla, D.~J.~Robinson and G.~Wormser,
Rev. Mod. Phys. \textbf{94}, no.1, 015003 (2022)
doi:10.1103/RevModPhys.94.015003
[arXiv:2101.08326 [hep-ex]].


\bibitem{Watanabe}
S.~Iguro and R.~Watanabe,
JHEP \textbf{08}, no.08, 006 (2020)
doi:10.1007/JHEP08(2020)006
[arXiv:2004.10208 [hep-ph]].

\bibitem{Bordonea}
M.~Bordone, M.~Jung and D.~van Dyk,
Eur. Phys. J. C \textbf{80}, no.2, 74 (2020)
doi:10.1140/epjc/s10052-020-7616-4
[arXiv:1908.09398 [hep-ph]].


\bibitem{Bordoneb}
M.~Bordone, N.~Gubernari, D.~van Dyk and M.~Jung,
Eur. Phys. J. C \textbf{80}, no.4, 347 (2020)
doi:10.1140/epjc/s10052-020-7850-9
[arXiv:1912.09335 [hep-ph]].





\bibitem{Workman:2022ynf}
R.~L.~Workman \textit{et al.} [Particle Data Group],
PTEP \textbf{2022}, 083C01 (2022)
doi:10.1093/ptep/ptac097

















\end{thebibliography}
\end{document}